\documentclass[conference]{IEEEtran}
\IEEEoverridecommandlockouts
\usepackage{url}
\usepackage{cite}
\usepackage{amsmath,amssymb,amsfonts}
\usepackage{algorithmic}
\usepackage{graphicx}
\usepackage{textcomp}
\usepackage{color,soul}
\usepackage{multirow}
\usepackage[table,xcdraw]{xcolor}
\def\BibTeX{{\rm B\kern-.05em{\sc i\kern-.025em b}\kern-.08em
    T\kern-.1667em\lower.7ex\hbox{E}\kern-.125emX}}
\begin{document}

\title{Energy consumption assessment of a Virtual Reality Remote Rendering application over 5G networks\\
% {\footnotesize \textsuperscript{*}Note: Sub-titles are not captured for https://ieeexplore.ieee.org  and should not be used}
}

\author{\IEEEauthorblockN{Roberto Viola}
\IEEEauthorblockA{\textit{Fundación Vicomtech} \\
\textit{Basque Research and Technology Alliance}\\
San Sebastián, Spain \\
rviola@vicomtech.org}
\\
\IEEEauthorblockN{Minh Nguyen, Alexander Zoubarev,\\Alexander Futasz, Louay Bassbouss}
\IEEEauthorblockA{\textit{Fraunhofer FOKUS} \\
Berlin, Germany\\
minh.nguyen@fokus.fraunhofer.de}
\and
\IEEEauthorblockN{Mikel Irazola, José Ramón Juárez}
\IEEEauthorblockA{\textit{Ikerlan Technology Research Center} \\
\textit{Basque Research and Technology Alliance} \\
Arrasate/Mondragón, Spain \\
mirazola@ikerlan.es}
\\
\IEEEauthorblockN{Amr A. AbdelNabi, Javier Fernández Hidalgo}
\IEEEauthorblockA{\textit{i2CAT Foundation}\\
Barcelona, Spain \\
amr.abdelnabi@i2cat.net}
}

\maketitle

\begin{abstract}
This paper investigates the energy implications of remote rendering for Virtual Reality (VR) applications within a real 5G testbed. Remote rendering enables lightweight devices to access high-performance graphical content by offloading computationally intensive tasks to Cloud-native Network Functions (CNFs) running on remote servers. However, this approach raises concerns regarding energy consumption across the various network components involved, including the remote computing node, the 5G Core, the Radio Access Network (RAN), and the User Equipment (UE). This work proposes and evaluates two complementary energy monitoring solutions, one hardware-based and one software-based, to measure energy consumption at different system levels. A VR remote renderer, deployed as CNF and leveraging the Media over QUIC (MoQ) protocol, is used as test case for assessing its energy footprint under different multimedia and network configurations. The results provide critical insights into the trade-off between energy consumption and performance of a real-world VR application running in a 5G environment.
\end{abstract}

\begin{IEEEkeywords}
Edge computing, Energy monitoring, Extended Reality, Network testbed, Performance analysis.
\end{IEEEkeywords}

\section{Introduction}

The usage of Extended Reality (XR) technologies, including Augmented Reality (AR) and Virtual Reality (VR), is accelerating across multiple sectors, driven by its potential to deliver immersive and interactive experiences. However, XR applications impose significant demands on network and computational resources, often exceeding the capabilities of mobile and wearable devices \cite{siriwardhana2021survey}. Thus, remote rendering has emerged as an alternative, wherein the computational workload is executed on a remote server, and only the rendered output is transmitted to the end device \cite{yeregui2024edge}. This approach leverages Cloud-native Network Functions (CNFs) to enable flexible and scalable service deployment on the remote server \cite{deng2024cloud}.

While remote rendering can effectively enable XR on resource-constrained devices, it introduces new challenges concerning the energy efficiency of the system \cite{chandio2024scoping}. The continuous processing of the graphical content through CNFs on the remote server and the streaming of the result through the communication infrastructures, including the 5G Core, the Radio Access Network (RAN) and the User Equipment (UE), can result in substantial energy consumption. As energy efficiency becomes a key concern for sustainable 6G networks, there is a pressing need to systematically assess the energy costs associated with the network services \cite{kerboeuf2024design}.

Existing literature offers limited insight into the energy footprint of XR rendering in operational 5G environments, as the analyses are usually limited to the XR application \cite{kattakinda2024towards} or the network \cite{williams2022energy}.
% Moreover, comprehensive tools and methodologies to monitor energy consumption across heterogeneous components, spanning hardware and software domains, are scarce.
To address this gap, this work presents a dual framework approach for energy monitoring within a real 5G testbed and focusing on remote rendering of VR multimedia content. The contributions are summarised as follows:
\begin{itemize}
    \item Implementation of two complementary energy monitoring frameworks in a real 5G testbed. The first employs external measurement devices to obtain accurate energy profiles of hardware components, while the second relies on software instrumentation to assess the energy usage of individual services and processes.
    \item Deployment of a VR remote renderer as CNF, leveraging beyond state-of-the-art Media over QUIC (MoQ) protocol for low-latency multimedia streaming from the edge to lightweight web clients over 5G infrastructure.
    \item In-depth comparative analysis of hardware and software energy monitoring approaches under various multimedia and network configurations. This evaluation highlights the advantages and limitations of each one, offering practical recommendations for future sustainable XR deployments.
\end{itemize}

\section{Dual energy monitoring framework}

% \hl{Image with general architecture, it could be generated by generalizing Fig.1 (If Fraunhofer provides it, I can edit and generate it) to include both hardware and software. This image could be generic, no mentioning any specific framework employed. The main section should explain why they are complementary. Then, in the subsection the actual implementation can be described.}

\begin{figure}
    \centering
    \includegraphics[width=\linewidth]{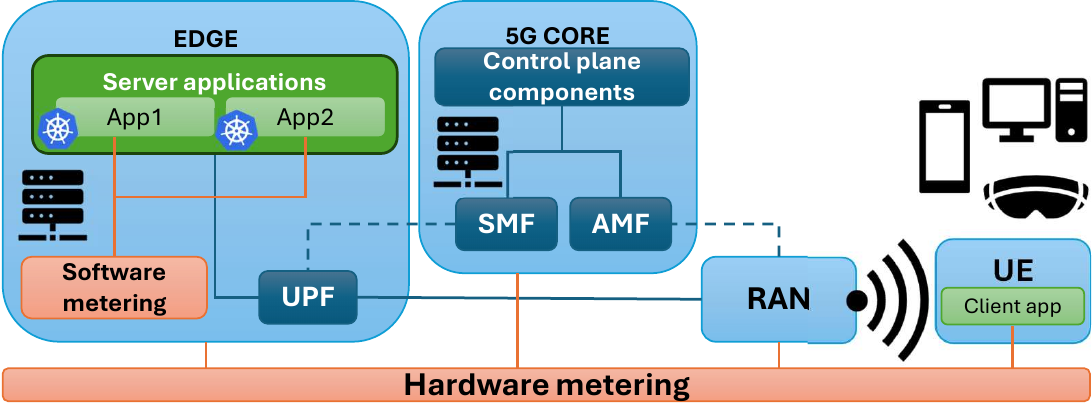}
    \caption{Dual energy monitoring architecture.}
    \label{fig:energy_architecture}
\end{figure}

To accurately characterise the energy footprint of network applications, Figure \ref{fig:energy_architecture} illustrates how hardware and software monitoring approaches can be combined to leverage complementary aspects of energy measurement. Hardware-level monitoring, typically using external power meters, provides highly accurate measurements of the total energy consumed by physical components such as UE, RAN and servers. These values serve as ground-truth references, capturing system-wide consumption, including energy flows that are inaccessible to the operating system (OS). However, hardware monitoring lacks the resolution needed to attribute energy use to individual services. To address this, software-based tools collect telemetry from platform-specific sources, such as CPU counters and GPU drivers, to estimate the energy consumed by specific processes. Although these estimates are less precise, they offer process-level granularity.

The integration of both approaches yields a comprehensive view, as hardware ensures measurement fidelity, while software enables detailed attribution. This dual strategy is essential for evaluating the trade-off between energy and performance, and guiding the optimisation of applications. With its generic design, this dual approach is applicable not only to VR but to any CNF-enabled application. In the next subsections, the implementation details of the two complementary approaches are further discussed.
% The integration of both approaches yields a comprehensive view, as hardware ensures measurement fidelity, while software enables detailed attribution. This dual strategy is essential for evaluating the trade-off between energy and performance, and guiding the optimisation of applications such as VR remote rendering. In the next subsections, the implementation details of the two complementary approaches are further discussed.

\subsection{Hardware-based energy monitoring}
% \hl{It seems a copy/paste of the final report, it is necessary re-elaborate it and focus to the objectives of the paper: no mention to unnecessary things like ffmpeg, exoplayer, etc.}

% \begin{figure}
%     \centering
%     \includegraphics[width=\linewidth]{figures/fraunhofer_measuring_architecture.png}
%     \caption{Overview of the hardware energy consumption}
%     \label{fig:fraunhofer_measuring_architecture}
% \end{figure}
% This section describes the architecture of the measuring system used to evaluate the energy consumption of 5G networks.

\begin{figure}
    \centering
    \includegraphics[width=0.80\linewidth]{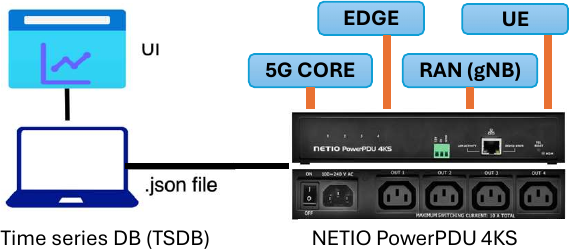}
    \caption{Hardware-based energy monitoring.}
    \label{fig:hw_monitor}
\end{figure}

Figure \ref{fig:hw_monitor} shows the implemented hardware-based solution to monitor the 5G network. The NETIO PowerPDU 4KS\footnote{\url{https://www.netio-products.com/en/device/powerpdu-4ks}} is employed as a smart power socket (power meter) to provide energy to all the physical components, i.e., edge server, 5G Core, RAN and UE. This allows the NETIO device to perform energy-related measurements, where the results are accessible through JSON content over HTTP.
These are visualised through a web dashboard in real-time and later stored in a time series database (TSDB) for more insightful analysis.
% These are later stored in a time series database (TSDB) and visualized through a custom web application dashboard. \hl{Fraunhofer: some revenant details to add? ==> Minh: Updated text in blue color.}
Energy consumption measurements are assessed for each of the physical nodes, while the overall End-to-End (E2E) consumption simply consists of their sum.

It is worth mentioning that in the case of a mobile UE, this solution works only if the UE is connected to the smart power socket through the charging cable and its battery is removed. If the battery cannot be removed, then the best solution is to use the device with a full charge, such that the energy consumed during the performed test is exactly the measured one.

\subsection{Software-based energy monitoring.}

\begin{figure}
    \centering
    \includegraphics[width=\linewidth]{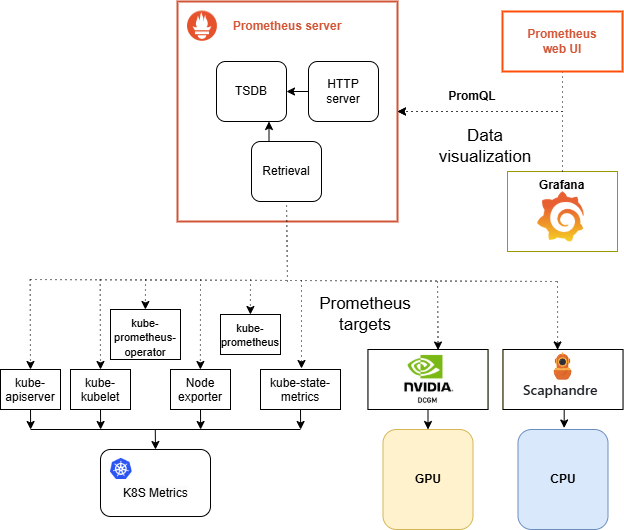}
    % \caption{Architecture of the software measurement system}
    \caption{Software-based energy monitoring}
    \label{fig:ikerlan_architecture}
\end{figure}

Figure \ref{fig:ikerlan_architecture} illustrates the software measurement system based on the Kubernetes ecosystem. The monitored applications at edge infrastructure are containerised, i.e., CNFs, and deployed on the cluster. The system leverages Prometheus\footnote{\url{https://prometheus.io/}} TSDB to collect metrics by scraping several targets:
% The solution is designed to monitor energy and power consumption of VNFs deployed at 5G edge.

% At the core of the system is the \textit{Prometheus server}, which is responsible for scraping metrics
% % from various configured targets
% , storing them in its TSDB and make them accessible via HTTP.
% Prometheus retrieves data from diverse sources, such as Kubernetes components, node-level exporters, and external hardware instrumentation tools. Queries are performed using PromQL, i.e., Prometheus native query language.
% For visualization and analysis, \textit{Grafana} is used to construct customizable dashboards. Grafana interfaces with Prometheus via PromQL, allowing users to analyse resource consumption patterns, detect anomalies, and infer energy behaviour trends.
% A wide range of scraping targets are employed:

\begin{itemize}
    \item Kubernetes-native exporters, including kube-apiserver, kubelet and node-exporter, expose metrics related to node resource consumption and pod lifecycle events.
    \item kube-state-metrics exports Kubernetes object state metrics such as deployments, replicas and namespaces.
    \item \textbf{CPU metrics} are gathered using Scaphandre\footnote{\url{https://github.com/hubblo-org/scaphandre}}, a power and CPU telemetry exporter. It provides insights into CPU usage, energy footprint, and per-process attribution.
    \item \textbf{GPU metrics} are collected using NVIDIA DCGM exporter\footnote{\url{https://github.com/NVIDIA/dcgm-exporter}}, offering fine-grained telemetry including GPU utilisation, temperature, and power consumption.
    % \item The kube-prometheus-operator is employed to manage the Prometheus lifecycle, automate service discovery, and configure alerting and recording rules.
\end{itemize}

The information coming from these sources allows to assess the energy consumption of each individual CNF. Additionally, Grafana\footnote{\url{https://grafana.com/}} allows to visualize the collected metrics.

\section{Experimental setup and case study}

This section presents the experimental environment used to validate the proposed dual energy monitoring framework. In particular, the 5G testbed, where the monitoring solutions are integrated, and the remote renderer, as an energy-demanding process to be monitored, are described.

\subsection{5G network testbed}
% \textcolor{blue}{---Amr and Javier finished the text and the figure}
\begin{figure}
    \centering
    \includegraphics[width=\linewidth]{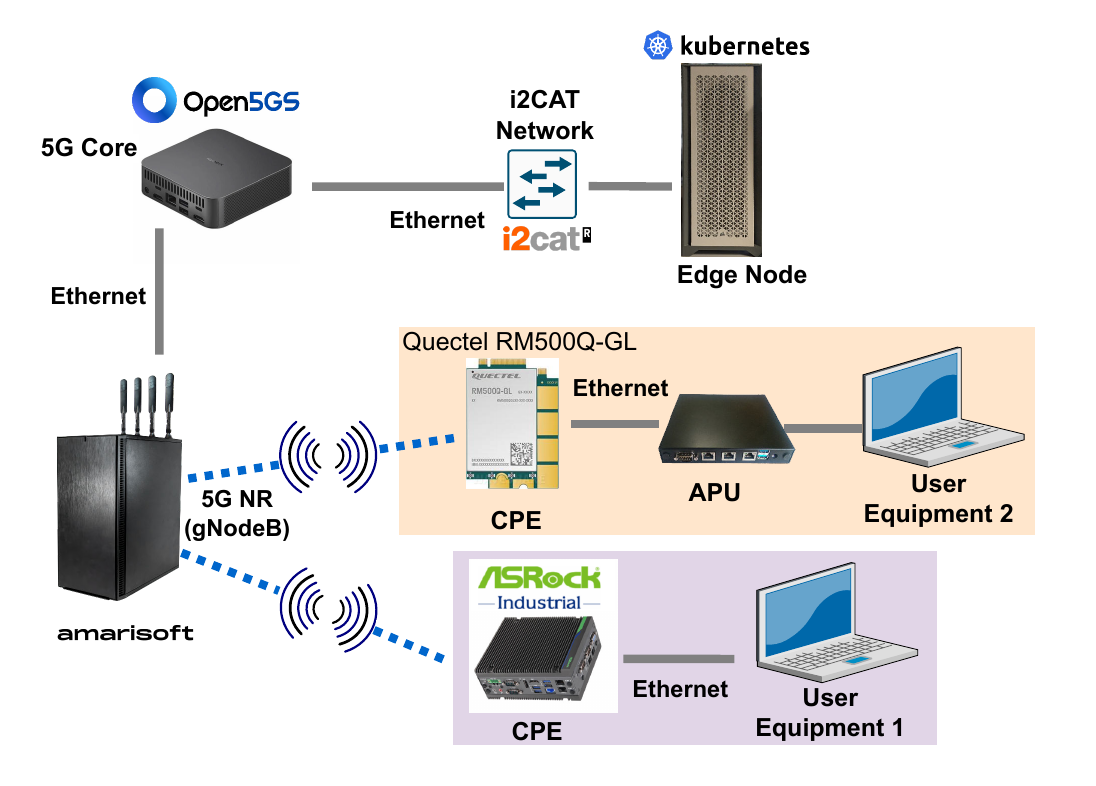}
    \caption{Architecture of the 5G network testbed.}
    % \caption{Architectural Layout of the 5G Standalone Testbed Integrating Radio Access, Core Network, and edge Computing for XR and Video Streaming Workloads.}
    \label{fig:5gnetwork}
\end{figure}

The 5G network, shown in Figure \ref{fig:5gnetwork}, is provided by 6G-XR project.
% and based on a typical standalone 5G architecture comprising RAN, Core and the edge node.
An Amarisoft Callbox Ultimate\footnote{\url{https://www.amarisoft.com/test-and-measurement/device-testing/device-products/amari-callbox-ultimate}} is employed as gNodeB, operating in the n77 \mbox{mid-band} \mbox{(3.3–4.2 GHz)}. It is configured in Time Division Duplexing (TDD) mode using a slot pattern that favours downlink transmission, which aligns with the high-throughput requirements of XR streaming traffic. The channel bandwidth can be configured with 40 or \mbox{100 MHz}.
The gNodeB is interfaced to the 5G Core based on Open5GS\footnote{\url{https://open5gs.org/}} through a direct Ethernet connection to ensure a low-latency communication with the gNodeB. 
% and deterministic transport behaviour. 

% The compute segment is further divided into Cloud and edge resources. For the scope of this work, the focus is placed on the edge node.
The edge node is implemented on a tower PC equipped with an AMD Ryzen Threadripper 3970X processor, featuring 32 cores operating at a base frequency of 3.70 GHz. The system is provisioned with 16 GB RAM and an NVIDIA RTX 3060 Ti GPU. The OS is Ubuntu Linux, which hosts a single-node Kubernetes cluster in which the remote rendering system and the software monitoring solution are deployed.

Two UEs are connected to the gNodeB to receive the content from the edge node. The first one consists of a laptop using an Asrock 5G Customer Premises Equipment (CPE)\footnote{\url{https://www.asrockind.com/en-gb/iEP-7020E}}, while the second one comprises a laptop
% with
interfaced via an Access Processing Unit (APU) to
a Quectel RM500Q-GL 5G modem\footnote{\url{https://www.quectel.com/product/5g-rm500q-gl/}}.
% , enabling PCIe-based 5G access.
Both UEs establish independent radio sessions with the gNodeB to allow isolated performance monitoring.
% and dynamic traffic analysis.

% This architecture enables comprehensive experimentation with media delivery for immersive applications, emphasizing end-to-end latency, jitter, bandwidth adaptation, and remote rendering responsiveness. The TDD configuration favouring downlink ensures optimal conditions for testing high-resolution video streams and complex XR scenes rendered at the edge and streamed to the UEs over 5G.

\subsection{VR remote rendering application}

\begin{figure}
    \centering
    \includegraphics[width=\linewidth]{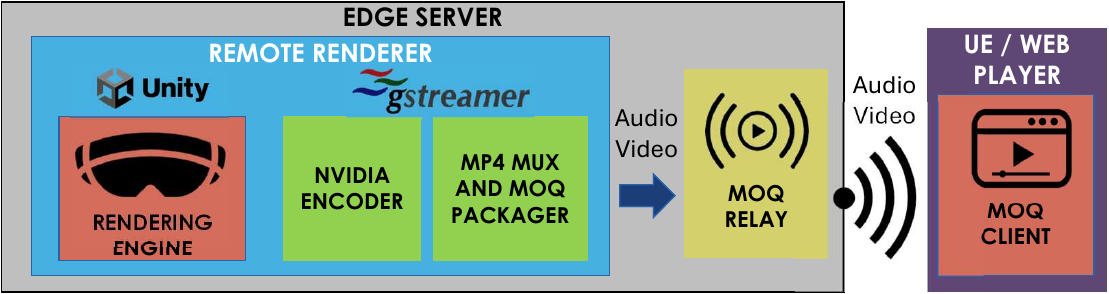}
    % \caption{Architecture of the software measurement system}
    \caption{Remote renderer with MoQ transmission to Web player.}
    \label{fig:renderer}
\end{figure}

Figure \ref{fig:renderer} shows the remote rendering system, implemented within the 6G-XR project \cite{mejías2025streaming}, and deployed as a CNF on the edge node to test the dual energy monitoring framework. It is designed to deliver 360°-rendered VR content to lightweight clients via the novel MoQ protocol. Its main components are:

\begin{itemize}
    \item \textbf{Remote Renderer:} It renders the VR content (i.e., VR scene and 3D objects) in a 360° video, then encodes it for streaming. It uses Unity and GStreamer frameworks, backed by GPU acceleration at the edge node.
    % running their processes on the GPU at the edge node.
    \item \textbf{MoQ Relay:} It acts as an intermediary that receives the MoQ media stream generated by the Remote Renderer and forwards it to the Web player. MoQ ensures low-latency and reliable delivery from the edge node to the UE through a beyond state-of-the-art streaming protocol.
    \item \textbf{Web Player:} It consists of a video player that receives and decodes the MoQ stream from the MoQ Relay. Since it is based on Web technologies, it runs in a common UE provided with a Web browser, such as a laptop.
\end{itemize}

\subsection{Testing scenarios}

The 5G testbed and the remote rendering application are configured in different ways to define the various scenarios:
\begin{itemize}
    \item \textbf{Idle\_UEs} (no rendering): the two UEs are connected to the 5G network, but the media streaming is stopped.
    \item \textbf{1\_active\_UE}: one UE receives and plays the 360° video generated by the Remote Renderer.
    % The test is performed several times by adjusting the video encoding bitrate, including 10 Mbps (low bitrate), 25 Mbps (moderate bitrate), and 40 Mbps (high bitrate) with H264 encoder.
    \item \textbf{2\_active\_UEs}: two UEs receive and play the 360° video generated by the Remote Renderer.
    % encoded at 40 Mbps. Two tests are performed with different network bandwidths at the RAN: 100 MHz, and 40 MHz.
\end{itemize}

Moreover, when one or two UEs are active, the 360° video stream from the Remote Renderer to the Web Player can be configured in one of the following ways:
\begin{itemize}
    \item \textbf{360\_10M}: MoQ streaming of 360° video encoded at 10 Mbps with NVIDIA H.264 encoder.
    \item \textbf{360\_25M}: MoQ streaming of 360° video encoded at 25 Mbps with NVIDIA H.264 encoder.
    \item \textbf{360\_40M}: MoQ streaming of 360° video encoded at 40 Mbps with NVIDIA H.264 encoder.
\end{itemize}

\section{Results and Discussion}

This section analyses the experimental results obtained from the dual energy monitoring framework integrated on the 5G testbed to monitor the Remote Renderer. Two complementary aspects are evaluated. First, a comparison of hardware and software techniques is presented. Then, the E2E energy behaviour across the overall 5G architecture is assessed.

\subsection{Comparison of hardware and software solutions}

The goal of comparing hardware and software solutions is to assess the accuracy of each of them under controlled workloads. To ensure a fair comparison, 5G Core, RAN, UE are not considered as the software solution is available only at the Kubernetes-enabled edge node.
This comparison includes only the \textit{Idle\_UEs} and \textit{1\_active\_UE} scenarios, as the remote renderer generates a single MoQ broadcast stream, meaning that the workload at the edge node does not increase with additional UEs.
Furthermore, measurements from both hardware and software solutions are synchronised using relative time, i.e., time since the experiment began.

Three power consumption metrics are compared: total hardware-level, host OS-level using the software monitoring, and container-level one relying on Kubernetes.
In the case of container-level, the power consumption of the pod executing the remote renderer is considered.

\begin{table}
\centering
\caption{Power consumption of edge node in \textit{Idle\_UEs} scenario.}
\label{tab:idle_power}
\begin{tabular}{|c|c|c|}
\hline
\multicolumn{3}{|c|}{\textbf{Average power consumption}} \\ \hline
\textbf{Hardware} & \textbf{Host OS} & \textbf{Remote Renderer} \\ \hline
133.60 W & 106.85 W & 20.68 W \\ \hline
\end{tabular}
\end{table}

Table \ref{tab:idle_power} presents the average power consumption in the \textit{Idle\_UEs} scenario, where the result at the host OS clearly shows an underestimation of -20\% compared to the hardware one. This systematic underestimation aligns with findings from energy measurement studies in the literature \cite{cao2020towards}. This is explained by underlying hardware activity that cannot be measured by the OS, e.g., fan consumption. While considering the Remote Renderer container (Kubernetes pod), its consumption appears to be only 19\% of the host OS consumption. It is worth noting that pod-level telemetry might present limitations, as it captures only container-level metrics and lacks visibility of system-wide components such as kernel threads or background processes that are actually used by Kubernetes and the OS to run the Remote Renderer container.

\begin{table}
\centering
\caption{Power consumption of edge node in \textit{1\_active\_UE} scenario for each multimedia configuration: absolute values and increases relative to \textit{Idle\_UEs} scenario.}
\label{tab:power_increase}
\begin{tabular}{|c|c|c|c|}
\hline
\multirow{3}{*}{\begin{tabular}[c]{@{}c@{}}\textbf{Multimedia}\\ \textbf{configuration}\end{tabular}} 
& \multicolumn{3}{|c|}{\textbf{Average power consumption}} \\ \cline{2-4}
& \textbf{Hardware} & \textbf{Host OS} & \textbf{Remote Renderer} \\
& \textbf{(\% increase)} & \textbf{(\% increase)} & \textbf{(\% increase)} \\ \hline
\multirow{2}{*}{360\_10M} & 223.43 W & 166.88 W & 64.49 W \\
& (+67.24\%)  & (+56.18\%)  & (+211.87\%) \\ \hline
\multirow{2}{*}{360\_25M} & 223.75 W &  166.94 W & 64.99 W \\
& (+67.48\%) & (+56.24\%)  & (+214.29\%) \\ \hline
\multirow{2}{*}{360\_40M} & 225.85 W  & 166.66 W  & 64.60 W \\
& (+69.05\%) & (+55.98\%) & (+212.40\%) \\ \hline
\end{tabular}
\end{table}

In Table \ref{tab:power_increase}, measurements for \textit{1\_active\_UE} scenario are presented, while considering the three multimedia configurations. The increase in terms of percentage compared to \textit{Idle\_UEs} scenario are also presented. Both hardware and host OS measurements present a consistent increase of 67--69\% and 55--56\%, respectively, across the considered workloads. These results reflect the rising computational demand due to the remote renderer that is now processing and streaming the multimedia content.
In this case, the power consumption reported by the host OS is around 25\% lower than that measured by the hardware, indicating a consistent underestimation.
Nevertheless, host-level telemetry provides a consistent reflection of workload scaling, thus retaining significant value for comparative profiling, even if it may fall short in delivering precise estimations.

As expected, the results of the Remote Renderer container show the highest increase, as it is now consuming more than three times than in the idle state, corresponding to 38\% of the host OS consumption. It is worth noting that variations in the multimedia encoding bitrate result in only marginal differences. This behaviour is likely attributable to the use of the H.264 hardware encoder, whose high level of maturity and widespread adoption suggest extensive performance optimisation, which is consequently reflected in its energy efficiency.

\begin{figure}[t]
    \centering
    \includegraphics[width=\linewidth]{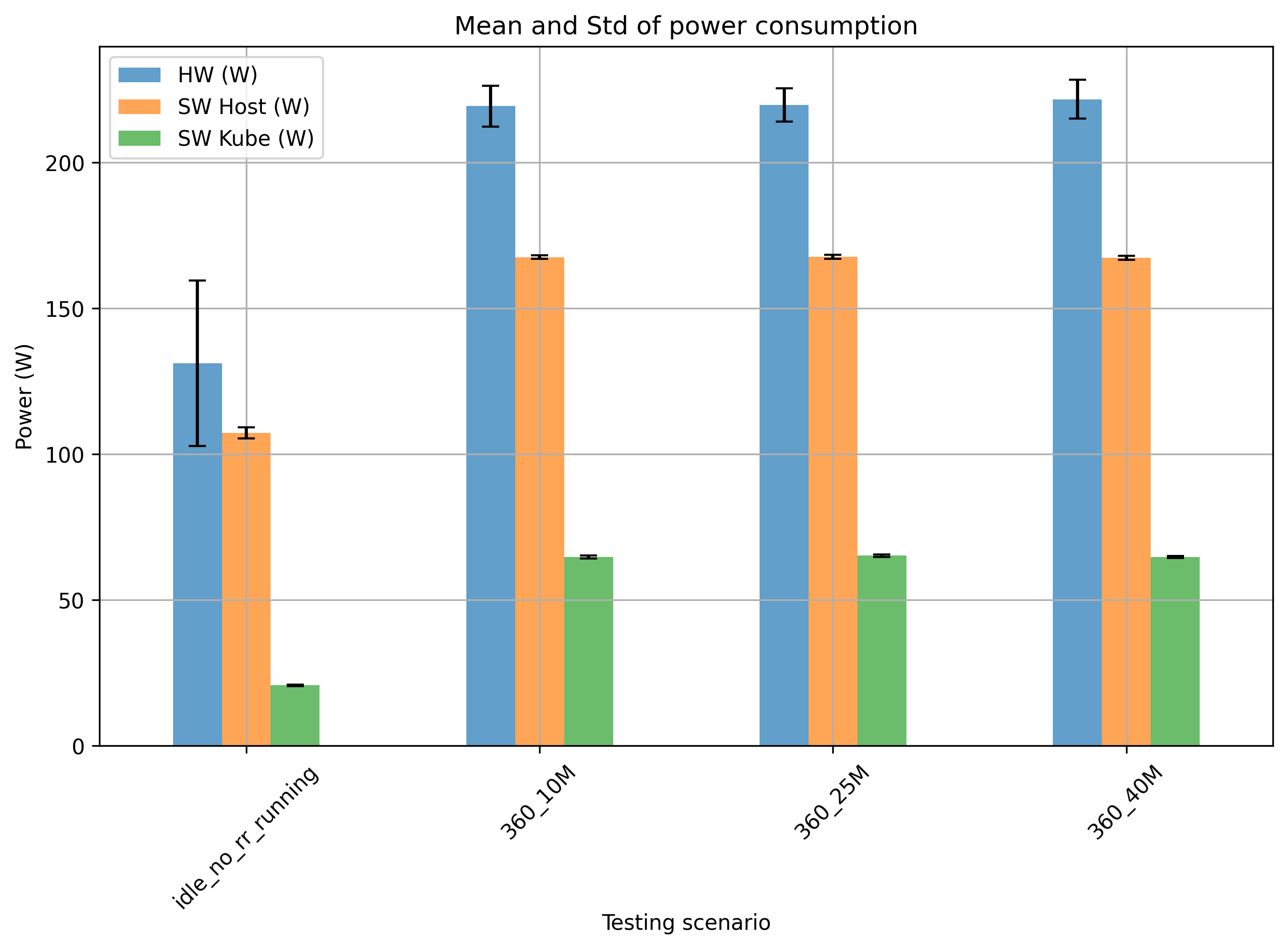}
    \caption{Average and standard deviation of power consumption across \textit{Idle\_UE} and \textit{1\_active\_UE} scenarios for each monitoring method. Three multimedia configurations are considered in \textit{1\_active\_UE} scenario.}
    \label{fig:mean_std_power}
\end{figure}

\begin{figure}[t]
    \centering
    \includegraphics[width=\linewidth]{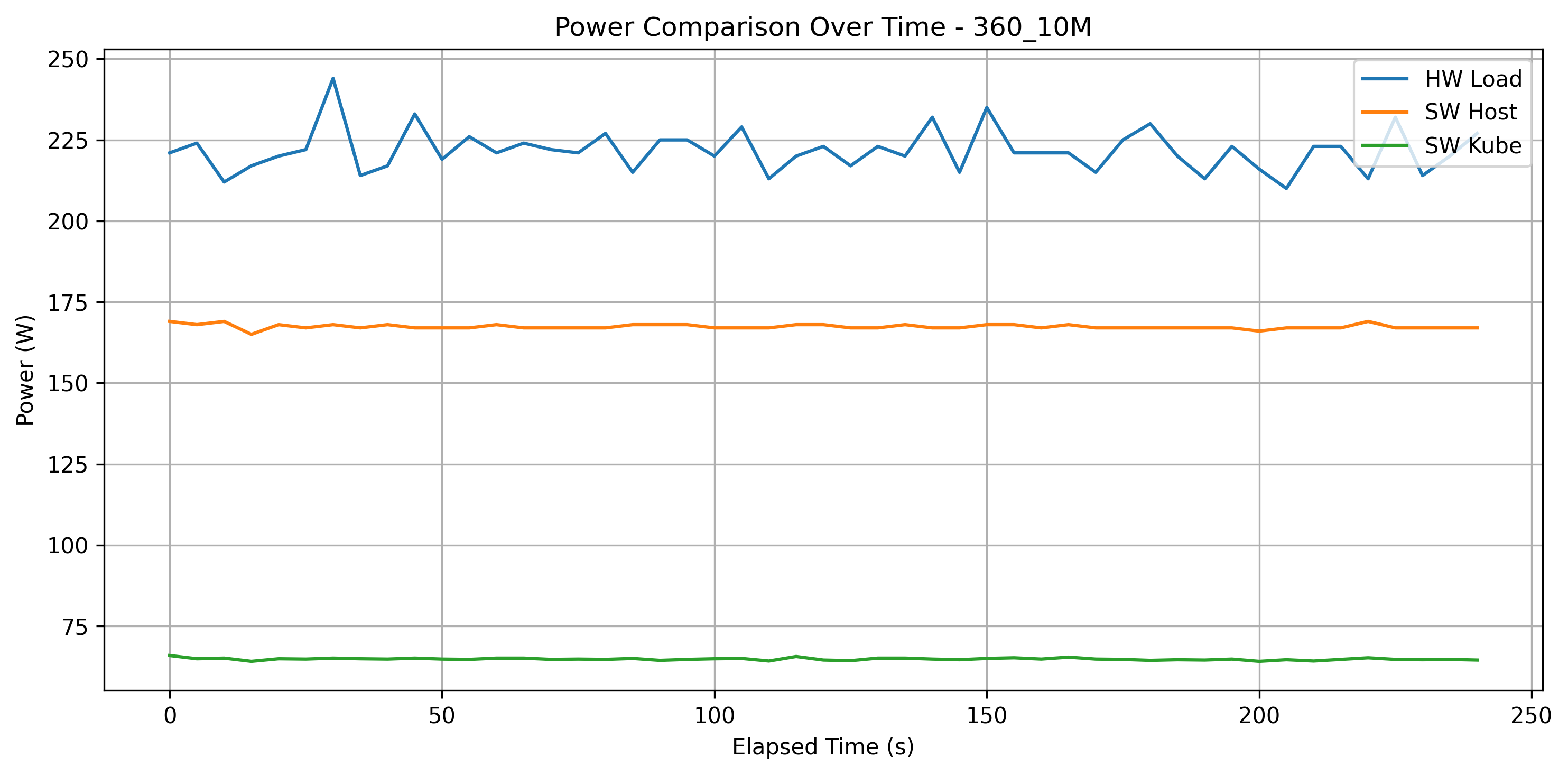}
    \caption{Power over time for the 360\_10M scenario. Software curves are significantly flatter than hardware.}
    \label{fig:360_10M_over_time}
\end{figure}

To further compare \textit{Idle\_UEs} and \textit{1\_active\_UE} scenarios, Figure \ref{fig:mean_std_power} presents a visual comparison of average and standard deviation for each monitoring method in such scenarios. The three multimedia configurations are considered when showing \textit{1\_active\_UE} scenario.
Hardware-based measurements consistently report both higher power usage and greater variability, while software-based telemetry, particularly at the Kubernetes level, shows reduced sensitivity to transient load dynamics.
This is also confirmed by the temporal patterns presented in Figure \ref{fig:360_10M_over_time}. The figure only shows the \textit{1\_active\_UE} scenario with \textit{360\_10M} multimedia configuration, even if similar results are obtained also with the other scenarios and configurations.
The hardware trace exhibits frequent fluctuations, indicating responsiveness to transient workloads and short-lived background processes. In contrast, both software traces, particularly the Kubernetes-level measurement, remain largely flat, suggesting that the software solution overlooks fine-grained fluctuations observed in the hardware traces. This lack of sensitivity results in missed peaks and inaccurate estimation of power variability.

To summarise, the results obtained from the comparison of hardware and software methods confirm that software-based metrics, particularly at the host level, are effective for capturing general usage trends, even though they underestimate total consumption. Container-level telemetry, while useful for per-process optimisation, might not represent the full power consumption behaviour.

\subsection{Analysis of E2E communication}

The goal of analysing the E2E communication is to provide insights into the energy footprint of the heterogeneous network components. In this case, only the hardware monitoring solution is employed as it can measure the power consumption at every node across the E2E communication.

% \subsubsection{Scenarios}
% The scenarios considered for multimedia transmission through a 5G network are the following:

% \begin{itemize}
%     \item \textbf{Idle UEs:} the UEs are connected to the 5G test network and no application in these UEs is running.
%     \item \textbf{One active UE:} One UE receives and plays a 360° video generated by the Remote Renderer, deployed at edge server, via MoQ protocol. The test is performed several times by adjusting the video encoding bitrate, including 10 Mbps (low bitrate), 25 Mbps (moderate bitrate), and 40 Mbps (high bitrate) with H264 encoder.
%     \item \textbf{Two active UEs:} two UEs receives and play a 360° video encoded at 40 Mbps. Two tests are performed with different network bandwidths at the RAN: 100 MHz, and 40 MHz.
    
% \end{itemize}

% \subsubsection{Analysis}
In the \textit{Idle\_UE} scenario, the power consumption is measured for two connected UEs that have no active applications, i.e., the VR rendering and streaming are not performed. The results in Table \ref{tab:idle} clearly show that the 5G Core consumes the least energy, while the gNodeB uses the most, followed by the edge server. The average powers of 5G Core, edge, and gNodeB are \mbox{5.2 W}, \mbox{129.6 W}, and \mbox{161.4 W}, respectively. The UEs, referring to the CPEs and their connected modems, require relatively low electricity, with only \mbox{15.9 W} for \mbox{UE 1} and \mbox{6.0 W} for \mbox{UE 2}.

% \usepackage{multirow}
% \begin{table}[t]
% \centering
% \caption{Energy consumption of 5G network with Idle UE scenario.}
% \label{tab:idle}
% \begin{tabular}{|c|ccccc|}
% \hline
% \multirow{2}{*}{\begin{tabular}[c]{@{}c@{}}Bandwidth\\ (MHz)\end{tabular}} & \multicolumn{5}{c|}{Average Power (W)}                                                                                  \\ \cline{2-6} 
%                                                                                   & \multicolumn{1}{c|}{Core} & \multicolumn{1}{c|}{gNodeB} & \multicolumn{1}{c|}{Edge}  & \multicolumn{1}{c|}{UE 1} & UE 2 \\ \hline
% 100                                                                               & \multicolumn{1}{c|}{5.2}  & \multicolumn{1}{c|}{161.4}  & \multicolumn{1}{c|}{129.6} & \multicolumn{1}{c|}{15.9} & 6.0  \\ \hline
% \end{tabular}
% \end{table}

\begin{table}
\scriptsize
\centering
\caption{Power consumption of each 5G network element in \textit{Idle\_UEs} scenario.}
\label{tab:idle}
\begin{tabular}{|c|c|c|c|c|c|}
\hline
\multirow{2}{*}{\begin{tabular}[c]{@{}c@{}}\textbf{Network}\\ \textbf{bandwidth}\end{tabular}} 
& \multicolumn{5}{c|}{\textbf{Average power consumption}} \\ \cline{2-6} 
& \textbf{Core} 
& \textbf{gNodeB} 
& \textbf{Edge} 
& \textbf{UE 1 (CPE)} 
& \textbf{UE 2 (CPE)} \\ \hline
100 MHz & 5.2 W & 161.4 W & 129.6 W & 15.9 W & 6.0 W \\ \hline
\end{tabular}
\end{table}

The results for the two scenarios involving active UEs are summarised in Table \ref{tab:active}, where different network and multimedia configurations are considered. The gNodeB is configured with a bandwidth of 40 or 100 MHz, while the 360° video stream is encoded at 10, 25 or 50 Mbps and delivered via MoQ protocol. 

\begin{table*}
\scriptsize
\centering
\caption{Power consumption of each 5G network element in \textit{1\_active\_UE} and \textit{2\_active\_UEs} scenarios.}
\label{tab:active}
\begin{tabular}{|c|c|c|c|c|c|c|c|}
\hline
\multirow{2}{*}{\textbf{Scenario}} & \multirow{2}{*}{\begin{tabular}[c]{@{}c@{}}\textbf{Network}\\ \textbf{bandwidth}\end{tabular}} & \multirow{2}{*}{\begin{tabular}[c]{@{}c@{}}\textbf{Multimedia}\\ \textbf{configuration}\end{tabular}} & \multicolumn{5}{c|}{\textbf{Average power consumption}} \\ \cline{4-8} 
& & & \textbf{Core} & \textbf{gNodeB} & \textbf{Edge} & \textbf{UE 1 (CPE)} & \textbf{UE 2 (CPE)} \\ \hline
\multirow{3}{*}{1\_active\_UE} & & 360\_10M & 5.75 W & 162.95 W & 219.52 W & 15.99 W & - \\ \cline{3-8}
& 100 MHz & 360\_25M & 5.97 W & 164.00 W & 220.15 W & 16.04 W & - \\ \cline{3-8}
&  & 360\_40M & 5.99 W & 164.38 W & 221.53 W & 16.07 W & - \\ \hline
\multirow{2}{*}{2\_active\_UEs} & 40 MHz & 360\_40M & 6.42 W & 156.88 W & 222.24 W & 16.03 W & 7.06 W \\ \cline{2-8}
& 100 MHz & 360\_40M & 6.39 W & 167.13 W & 221.32 W & 16.03 W & 7.46 W \\ \hline
\end{tabular}
\end{table*}

In \textit{1\_active\_UE} scenario and employing a network bandwidth of \mbox{100 MHz}, the average power consumption of the UE itself remains relatively stable, ranging from \mbox{15.99 W} at \mbox{10 Mbps} to \mbox{16.07 W} at \mbox{40 Mbps}. This indicates that the energy consumption of the UE does not significantly increase with higher bitrates. The edge component is the one that consumes the most power among the components, starting at \mbox{219.52 W} at \mbox{10 Mbps} and increasing to \mbox{221.53 W} at \mbox{40 Mbps}. It can be seen that increasing the bitrate of the video (from 10 to \mbox{40 Mbps}) also increases the energy consumption but not significantly (i.e., increase by less than 1\%). The edge consumption reflects its role in processing and delivering content to the UE.
% Similarly to what was explained in the previous section, this limited increase in power consumption at the UE and edge when increasing the encoding processing is likely attributable to the use of the H.264 codec. The implementations of the encoder at the Remote Renderer (edge) and the decoder at the Web Player (UE) might have a high level of maturity, which is reflected in their energy efficiency.
% Similarly to what was explained in the previous section,
Consistent with our previous observations, this limited increase in power consumption at the edge when increasing the encoding processing is likely attributable to the use of the H.264 codec. The mature GPU encoder implementation at the Remote Renderer (edge) is reflected in its energy efficiency.

A marginal increase is also seen in the consumption of 5G Core, which suggests that its energy demands are primarily driven by the overhead of managing the connection rather than the amount of data being processed. The gNodeB shows a slight increase in power consumption from \mbox{162.95 W} at \mbox{10 Mbps} to \mbox{164.38 W} at \mbox{40 Mbps}. Even if this consists of a less than 1\% of increase, it reflects the additional resources required to manage higher data rates and maintain the throughput necessary for video streaming. The gNodeB consumption is primarily linked to its role in signal processing and the transmission of data to the UE. The relatively stable power usage indicates that while the gNodeB needs to allocate more resources for higher bitrates, it does so efficiently, with only a minor increase in power draw.

\begin{figure}
    \centering
    \includegraphics[width=\linewidth]{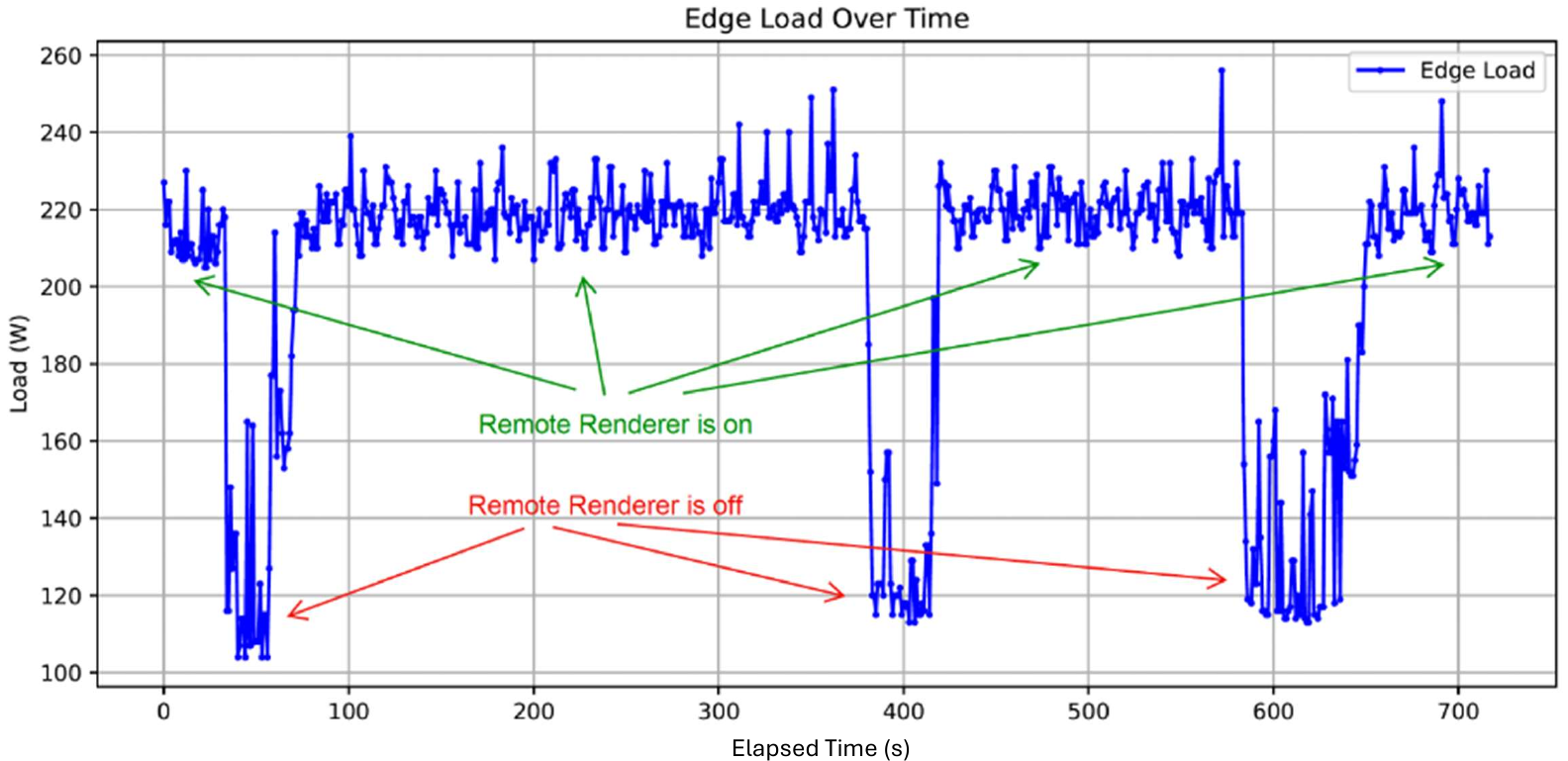}
    \caption{Power consumption of the edge when the Remote Renderer is running (around 220 W) or not (around 120 W).}
    \label{fig:renderer_on_of}
\end{figure}

With the same network bandwidth of \mbox{100 MHz}, but considering the \textit{2\_active\_UEs} scenario instead, the power consumption for the 5G Core and gNodeB rises to \mbox{6.39 W} (+6.7\%) and \mbox{167.13 W} (+1.7\%), respectively, with two UEs streaming at \mbox{40 Mbps}.
Meanwhile, the power consumption of the edge and UE 1 is relatively unchanged compared to when one UE is active. This is explainable as the Remote Renderer at the edge creates only one broadcast stream via MoQ, i.e., it creates only one stream that is shared among all the UEs that consume it. It means that the edge and UE processing, and thus the energy consumption, remain the same as the \textit{1\_active\_UE} scenario.
When comparing with the \textit{idle\_UEs} scenario, there is an increase of 11\% and 25\% in the energy consumption of the 5G Core and UE 2, while the results for gNodeB and UE 1 are almost unchanged. In comparison, the edge server consumes significantly more energy when the Remote Renderer is activated. Its energy consumption jumps from \mbox{129.6 W} to \mbox{221.32 W} (i.e., by +71\%), which reveals that the remote rendering process is a heavy task that should be shut down whenever possible. Figure \ref{fig:renderer_on_of} shows the difference in the energy consumption of the edge when the Remote Renderer is running or not.

Finally, the shift from 100 MHz to 40 MHz bandwidth in the \textit{2\_active\_UEs} scenario does not lead to a significant reduction in energy consumption, suggesting that the benefits of lower bandwidth may not outweigh the energy demands of maintaining multiple active streams.

% The key findings and suggestions based on the above analysis can be drawn as follows.
To summarise, the results obtained from the analysis of the E2E communication suggest that the gNodeB and edge components are the primary contributors to overall energy consumption, indicating that optimisation efforts should prioritise these elements. Notably, varying the video bitrate has minimal influence on the power usage of UE, suggesting that lowering the bitrate for energy savings may unnecessarily degrade video quality.
Furthermore, a substantial increase in power consumption is observed in the edge component when heavy processing, such as remote rendering, is enabled, emphasising the need for strategies to limit or optimise the usage of intensive processing tasks.

% \begin{itemize}
%     \item The gNodeB and the edge are the most energy-intensive components. It is necessary to focus on optimizing the power usage of these components for improved energy efficiency.
%     \item The impact of the video bitrate on the UE power is minimal. We do not recommend reducing the video bitrate to save energy because of unnecessary deductions in video quality.
%     \item Edge component shows a significant power consumption increase (by 71\%) when remote rendering is activated. We suggest developing and implementing strategies to limit remote rendering usage or optimize its efficiency.
%     \item Switching from 100 MHz to 40 MHz bandwidth does not significantly reduce energy consumption. It is not necessary to reduce the bandwidth to save energy.
% \end{itemize}

\section{Conclusions and Future Work}

% This work has presented a dual hardware and software energy monitoring framework. The combination of hardware and software measurements can guarantee higher accuracy of the hardware with higher granularity of the software.

% Hardware and software measurements have been evaluated in the context of a high-processing application, such as a VR remote rendering process deployed as a CNF on a real 5G testbed. The results have shown that software-based monitoring consistently underestimates actual energy consumption, especially at the container level. While host-level metrics have captured overall workload trends reasonably well, they have fallen short in reflecting short-term fluctuations and transient system activity that were instead visible in hardware measurements.
This work has presented a dual hardware and software energy monitoring framework evaluated using a VR remote rendering process, deployed as a CNF on a real 5G testbed. Combining the higher accuracy of hardware with the greater granularity of software provides a comprehensive view of energy consumption.

% The results show that software-based monitoring consistently underestimates actual energy consumption, especially at the container level, and fails to capture short-term fluctuations that were instead visible in hardware measurements. Despite these limitations, it remains useful for observing dynamic energy patterns and for providing more granularity, as they are particularly well-suited at the edge server to attribute energy usage to individual processes. Ultimately, for accurate energy profiling and full network node power accounting, hardware-based measurements remain essential.
The results show that software-based monitoring consistently underestimates actual energy consumption, especially at the container level, and fails to capture short-term fluctuations that were instead visible in hardware measurements. Despite these limitations, software monitoring remains useful for observing general usage trends and is particularly well-suited at the edge server to attribute energy usage to individual processes. Ultimately, for accurate energy profiling and full network node power accounting, hardware-based measurements remain essential.

Building upon these findings, future research should focus on improving software-based energy monitoring, particularly through the development of dedicated assessment tools for the 5G Core, RAN, and UE. Moreover, this work also provides the basis for developing orchestration strategies aimed at optimizing the energy consumption of future networks and services. Ultimately, the goal is to leverage the insights from this dual monitoring approach to develop a real-time system that can dynamically optimize resource allocation to balance performance with energy consumption.

% \begin{table}[htbp]
% \caption{Table Type Styles}
% \begin{center}
% \begin{tabular}{|c|c|c|c|}
% \hline
% \textbf{Table}&\multicolumn{3}{|c|}{\textbf{Table Column Head}} \\
% \cline{2-4} 
% \textbf{Head} & \textbf{\textit{Table column subhead}}& \textbf{\textit{Subhead}}& \textbf{\textit{Subhead}} \\
% \hline
% copy& More table copy$^{\mathrm{a}}$& &  \\
% \hline
% \multicolumn{4}{l}{$^{\mathrm{a}}$Sample of a Table footnote.}
% \end{tabular}
% \label{tab1}
% \end{center}
% \end{table}

% \begin{figure}[htbp]
% \centerline{\includegraphics{fig1.png}}
% \caption{Example of a figure caption.}
% \label{fig}
% \end{figure}

\section*{Acknowledgment}
% The preferred spelling of the word ``acknowledgment'' in America is without 
% an ``e'' after the ``g''. Avoid the stilted expression ``one of us (R. B. 
% G.) thanks $\ldots$''. Instead, try ``R. B. G. thanks$\ldots$''. Put sponsor 
% acknowledgments in the unnumbered footnote on the first page.
This research was supported by the SNS-JU Horizon Europe Research and Innovation programme, under Grant Agreement 101096838 for 6G-XR project.

\bibliographystyle{IEEEtran}
\bibliography{IEEEabrv,main.bib}

\end{document}